# Chromium-vacancy clusters in dilute bcc Fe-Cr alloys: an *ab initio* study

M.Yu. Lavrentiev[*], D. Nguyen-Manh, and S.L. Dudarev

*CCFE, UK Atomic Energy Authority, Culham Science Centre, Abingdon, Oxon, OX14 3DB, United Kingdom*

## Abstract

Using an *ab initio* approach, we explore the stability of small vacancy and vacancy-chromium clusters in dilute body-centred cubic Fe-Cr alloys. To explain experimental observations described in C.D. Hardie *et al.*, *J. Nucl. Mater.* **439**, 33 (2013) and showing the occurrence of Cr segregation in low-Cr alloys, we investigate if chromium can form stable bound configurations with vacancies in alloys with chromium concentration below the low-temperature chromium solubility limit of 10-11 at. %. We find that a single vacancy can attract up to four Cr atoms in the most energetically favourable cluster configuration. The binding energy of a cluster containing a single vacancy and from one to eight Cr atoms can be well described by a linear function of the number of chromium atoms in the second, third and fifth nearest neighbour coordination. The magnetic origin of the binding energy trend is confirmed by a correlation between the average value of the magnetic moment of a Cr atom and the binding energy. Similar trends are also found for di-vacancy-Cr clusters, confirming that they likely also characterise larger systems not yet accessible to *ab initio* calculations. The ratio of the binding energy to the number of Cr atoms increased more than twice in the di-vacancy case in comparison with a single vacancy case.

[*] Corresponding author. E-mail: Mikhail.Lavrentiev@ukaea.uk; Tel: +441235464329; Fax: +441235466435

**Introduction**

Iron-chromium alloys have attracted significant attention in the past decade because of a broad range of their applications in the fields of fusion and advanced fission [1]. It is well established that the enthalpy of mixing of Cr in body centred cubic (bcc) Fe is negative at small chromium concentrations, but it changes sign and becomes positive above 10-11 at.% chromium concentration in Fe-Cr alloys [2-4]. This change of sign is responsible for the occurrence of segregation and clustering of Cr in concentrated binary Fe-Cr alloys. In the low concentration limit, chromium atoms prefer to be surrounded by Fe atoms and separated from each other as much as possible; above 10-11 at.% clustering of Cr begins. This results in a rather unusual phase diagram of Fe-Cr alloys where the solubility of Cr in Fe in the dilute limit does not tend to zero at low temperatures [5,6].

Hence, clustering of chromium atoms in dilute alloys is not expected. However, recent atom probe tomography studies by Hardie *et al.* [7] have shown the surprising occurrence of large clusters of Cr atoms in Fe-5at.% Cr alloys irradiated with $Fe^+$ ions at 400°C. There are several possible explanations for the occurrence of such clusters, for example Cr enrichment at grain boundaries. There is also a possibility that vacancies created during irradiation form small stable complexes with Cr atoms, effectively causing clustering well below the concentration threshold corresponding to thermodynamic equilibrium. In recent studies by Barouh *et al.* [8] and Schuler *et al.* [9], it was found that vacancies in Fe could form stable complexes with interstitial oxygen and nitrogen. It is natural to explore whether vacancies could have a similar effect on the solute components of Fe alloys, resulting in the formation of relatively stable solute-vacancy clusters.

Energies and structures of pure vacancy clusters were investigated in several recent studies [10-12]. However, relatively little is known about vacancy-chromium clusters in iron-based alloys, which were investigated experimentally in [7]. Below, we summarise results of an *ab initio* investigation spanning a number of small vacancy-chromium clusters, to assess whether stable chromium-vacancy complexes could form in dilute Fe-Cr alloys. We investigate how the binding energy of vacancy-chromium clusters varies as a function of their size and geometrical configuration, in particular how it depends on the vacancy-vacancy and Cr-Cr separation. Magnetic properties of chromium and their relation to the strength of binding between chromium and vacancies are also investigated. We establish a linear fit relating the binding energy of a vacancy-Cr cluster to its geometrical configuration. This relationship is expected to provide a foundation for a cluster expansion based treatment of large clusters of vacancies and chromium atoms. Calculations performed here include configurations containing one or two vacancies and up to eight Cr atoms. This means that for the simulation cell containing 128 atoms, the total chromium concentration remains low, being no higher than 6.25 at. %, and hence well below the thermodynamic clustering threshold of 10-11 at. %.

The paper is organized as follows. First, we briefly introduce the methodology of *ab initio* calculations and define values of parameters controlling the simulations such as the cell size, the cutoff energy etc. Then, the total energy values computed for a single vacancy and small vacancy clusters in pure Fe are derived and compared to other studies as well as to experiment. Next, clusters consisting of a single vacancy and containing from one to eight Cr atoms in the first coordination shell are considered. We then proceed to studying clusters containing two vacancies and one or two chromium atoms. Finally, we discuss the simulation data and conclude.

**Computational details**

DFT calculations were performed using the projector augmented wave (PAW) method implemented in VASP [13,14]. Exchange and correlation were treated in the generalized gradient approximation GGA-PBE [15]. PAW-PBE potentials with semicore electrons were used for both Fe and Cr atoms. The plane-wave cutoff energy used in the calculations was 400 eV. The simulation box contained 4×4×4 unit cells of bcc Fe, i.e. 128 atoms for pure iron, and 127 or 126 atoms where we investigated clusters of vacancies and chromium. In order to check convergence as a function of system size, several calculations using 5×5×5 unit cell simulation boxes (250 atoms for pure Fe) were also performed. Energies were evaluated using the 4×4×4 Monkhorst-Pack mesh [16] of *k* points in the Brillouin zone. We performed full relaxation with respect to atomic coordinates and cell dimensions, and used spin-polarized electronic structure approach, to investigate magnetic properties of vacancy-chromium clusters and to estimate the amplitudes of displacements of atoms from their perfect crystallographic positions.

**Single vacancy and small vacancy clusters**

The formation energy of a single vacancy found in the calculations equals $E^f_{1vac} = 2.215$ eV. This value agrees with most of other *ab initio* studies, which give values in the range from 1.95 to 2.25 eV, depending on the size of the simulation cell, effects of volume relaxation and other computational details [13, 17-28]. Experimental observations often exhibit slightly lower values of vacancy formation energy namely $1.6 \pm 0.15$ eV [29], $1.7 \pm 0.1$ eV [30], with the highest known experimental vacancy formation enthalpy in the ferromagnetic phase being $2.0 \pm 0.2$ eV [31]. When discussing the relaxation of atomic positions around a vacancy, it is appropriate to define two distinct, although related, values: the formation volume and the relaxation volume. The former is defined as $\Omega^f_{1vac} = \Omega_0 + (\Omega_N - V(127Fe + 1Vac))$ [32], where $\Omega_N = N\Omega_0$, $N$ is the number of atoms in the simulation cell (128 in the present work) and $\Omega_0$ is volume per atom in a perfect lattice. The latter is an observable quantity that enters elasticity equations, and is defined as $\Omega^r_{1vac} = V(127Fe + 1Vac) - \Omega_N$ [33,34]. For the relaxation volume of a single vacancy in pure Fe we have found the value of -2.74 Å$^3$. This is in good agreement with the recent study by Murali *et al.* [28], where they found $\Omega^f_{1vac} = -2.92$ Å$^3$ adopting the same simulation cell size. Formation volume is usually presented as a fraction of the ideal volume for atom, $\Omega_0$. For it, we obtained $\Omega^f_{1vac}/\Omega_0 = 0.76$. This value lies inside a fairly broad range of values obtained in several previous calculations, namely 0.55 and 0.63 [17], 0.90 [18], 0.80 [19], and 0.65 [12]. Good agreement with literature values found in the case of a single vacancy makes it possible to extend out study to small vacancy complexes.

We have also investigated how the binding energy between two vacancies varies as a function of separation between them. The binding energy of a cluster of *m* vacancies was calculated according to the formula

$$E_b(mV) = m * E(127Fe + V) - (m-1) * E(128Fe) - E((128-m)Fe + mV) \qquad (1)$$

where *V* denotes a vacancy. A positive value of the binding energy, according to this definition, implies that a vacancy cluster is more favourable energetically that several individual single vacancies distributed randomly in the crystal. Taking into account the constraint associated with the size of the simulation box, distances up to the fifth nearest neighbour were investigated. Binding energies are given in Table 1, where they are compared with literature data. In agreement with previous results, the binding energy of a di-vacancy is the highest for the next nearest neighbour distance. Beyond the second nearest neighbour distance, the binding energy decreases substantially. At the third nearest neighbour separation and further, elastic interactions dominate binding between vacancies, and it is instructive to estimate this interaction. An expression for the energy of elastic interaction between two spherically symmetric defects in a cubic crystal (see, e.g., [35]) has the form:

$$E = -\frac{15 \Delta V_1 \Delta V_2 d}{8\pi \gamma^2 a^3} \frac{\Gamma}{(r/a)^3}, \quad (2)$$

$$\Gamma = \frac{3}{5} - \frac{(x^4 + y^4 + z^4)}{r^4}, \quad (3)$$

where $\Delta V_1$, $\Delta V_2$ are the relaxation volumes of the two interacting defects, $r = |\mathbf{r}|$ is the distance between them, $x, y, z$ are Cartesian coordinates of vector $\mathbf{r}$, $a$ is the unit cell size, $\gamma = 3(1-\nu)/(1+\nu)$, $\nu$ is the Poisson ratio of the material, $d = C_{11} - C_{12} - 2C_{44}$ is a measure of anisotropy of the material in terms of its elastic constants. Using for the unit cell size and the relaxation volume the values obtained in our calculations (2.824 Å and -2.74 Å$^3$, respectively), and for the Poisson ratio and elastic constants the experimental values ($\nu$ =0.291 [36], $C_{11}$=243.1 GPa, $C_{12}$=138.1 GPa, $C_{44}$=121.9 GPa [37]), we obtained for the dimensional factor $-\frac{15 \Delta V_1 \Delta V_2 d}{8\pi \gamma^2 a^3}$ the value of 63 meV. The dimensionless angular and distance factor $\frac{\Gamma}{(r/a)^3}$ for the third nearest neighbor equals $\frac{\sqrt{2}}{40} = 0.035$, and hence the strength of elastic interaction between vacancies at the third nearest neighbor separation is about only 2 meV (and repulsive, i.e. $E_b^{3NN}(2V) \approx -2$ meV). This value is an order of magnitude lower than the DFT result. To check that with increasing size of the simulation box the binding energy for the third nearest neighbour decreases, we performed several calculations using a larger box of 5×5×5 unit cells (250 atoms for pure iron). Results are given in Table 1 in parentheses and show that indeed the binding energy falls almost to zero for the third nearest neighbour distance between vacancies in agreement with the above estimate, whereas for the first and second nearest neighbours the values of the binding energy remain close to those obtained for the 4×4×4 simulation box.

Finally, we investigated the most compact tri-vacancy complex and two clusters of four vacancies adopting "tetragonal" and "square" configurations. These complexes are shown in

Figure 1. Results for the binding energies are also given in Table 1. The binding energies of tri-vacancy and "tetragonal" four-vacancy clusters agree well with recent calculations by Kandaskalov *et al.* [12]. They also indicate the high stability of the "tetragon"-shaped four-vacancy cluster. However, in the case of a "square" four-vacancy cluster, our calculated binding energy of 1.116 eV is substantially higher than that found in earlier calculations performed using interatomic potentials [10], tight-binding [11] and DFT [12] methods. The binding energy remains large (0.975 eV) for the "square" cluster also if the calculation is performed on a 5×5×5 unit cells simulation box. Increasing the size of the vacancy clusters leads to higher magnetic moment of neighbouring Fe atoms. In pure Fe, it is 2.22 $\mu_B$; around a single vacancy it is 2.43 $\mu_B$; for iron atoms that have two vacancies as nearest neighbours, magnetic moment approaches 2.62-2.65 $\mu_B$; and for the atoms with four nearest neighbour vacant sites (for example like those found in a "square" configuration) the magnetic moment is 2.81 $\mu_B$. This correlates well with results found in other computational studies, in particular in calculations showing the increase of Fe magnetic moment at the (100) surface up to 3 $\mu_B$ [38,39], as well as with recent experimental findings showing the increase of observed Fe magnetic moment following self-ion irradiation, which is believed to be associated with the production of a large number of vacancy clusters [40]. In Figure 2, we show values of magnetic moments for Fe atoms with one, two, and four nearest neighbour vacancies, as obtained for the case of a "square" vacancy cluster.

**Interaction of chromium atoms with a single vacancy**

We start from the investigation of interaction between a vacancy and a chromium atom. As in the case of two vacancies, distances up to the fifth nearest neighbour were studied. The resulting binding energies are given in Table 2. The strongest vacancy-Cr attraction is found in the nearest neighbour configuration, and the binding energy rapidly decreases as a function of separation between the vacancy and a chromium atom.

Next we consider the case of a single vacancy surrounded several Cr atoms. Similarly to the case of a vacancy cluster, the binding energy of a cluster containing *m* vacancies and *n* chromium atoms is defined as follows:

$$E_b(mV + nCr) = m*E(127Fe+V) + n*E(127Fe+Cr) - (m+n-1)*E(128Fe) \\ - E((128-m-n)Fe + mV + nCr) \tag{4}$$

A single vacancy can have between one and eight Cr atoms in its first coordination shell. Among the configurations involving from 2 to 6 chromium atoms, there are cases that are not related by symmetry operations. In total, there are 21 possible configurations, where for each configuration an *ab initio* calculation with full structural relaxation was performed. Figure 3 shows the binding energy of all such vacancy-chromium complexes as a function of the number of Cr atoms.

In the interval from one to four Cr atoms, there are configurations where the binding energy increases almost linearly as a function of the number of Cr neighbours, as shown by the red line in the Figure 3, increasing to the maximum value of 0.303 eV (this corresponds to approximately 0.076 eV per Cr atom). These configurations are shown in Figure 4. What they have in common is that Cr atoms are situated as far apart from each other as possible. For two Cr atoms, this corresponds to the fifth nearest neighbour coordination, for three and four atoms, they are the third nearest neighbours of each other. On the other hand, complexes with the lowest binding energy (the lowest binding energy is negative starting from three Cr atoms, indicating repulsion) are characterized by the second nearest neighbour coordination of chromium atoms, i.e. the closest interatomic distance that is possible given the constraint that Cr are in the first coordination shell of the vacancy. In the configuration where we have 8 Cr atoms near a vacancy, the binding energy is negative and equal to -0.888 eV. It is possible to approximate the binding energy of a single vacancy cluster containing from 1 to 8 Cr atoms by the following analytical linear expression

$$E_b(eV) = 0.15753 - 0.11411 * N(2NN) + 0.02193 * N(3NN) + 0.02737 * N(5NN) \qquad (5)$$

where *N(2NN), N(3NN), N(5NN)* are the numbers of the second, third and fifth nearest Cr-Cr neighbours in a configuration, respectively. This linear fit spans binding energies over a fairly broad range of ~1.2 eV and has the mean square error of only 0.029 eV. The fairly good agreement between the fit and the calculated DFT energies is illustrated in Figure 5.

The likely reason why simulations show a correlation between large Cr-Cr separations and positive values of the binding energy is the repulsive interaction between chromium atoms (in the Fe matrix) that have parallel magnetic moments. This fact was noted in earlier work (e.g., [3]), including our own work on bcc-fcc phase transitions in Fe-Cr [41], and can even give rise to magnetic non-collinearity at Fe-Cr interfaces and large chromium clusters in Fe [42,43]. The magnetic origin of Cr-Cr repulsion is also confirmed by the analysis of magnetic moments of chromium atoms. In Figure 6, the average value of the magnetic moment of Cr atoms is plotted versus the binding energy of a cluster computed for all the configurations. There is a clear correlation between the two quantities, where a higher value of the binding energy is correlated with a high absolute value of the average moment. Cr atoms closely packed around a vacancy tend to have lower magnetic moment as this minimizes repulsive Cr-Cr interactions, while for the well-separated Cr, large magnetic moments stimulate stronger Fe-Cr interaction, which overcomes Cr-Cr repulsion and results in a positive value of the binding energy. As a result, configurations with the maximum number of Cr atoms (four) as the third nearest neighbours, see Figure 4, are characterised by the highest binding energies. The influence of a vacancy on the magnetic moment of chromium atoms in its vicinity can be illustrated by the difference between Cr magnetic moment in Fe without a vacancy (1.693 $\mu_B$) and in the presence of a vacancy (1.936 $\mu_B$). In the case where there are eight Cr atoms around a vacancy, their magnetic moment is lower than the corresponding value of Cr moment in bulk Fe (1.574 $\mu_B$). Hence the presence of a vacancy increases moments of both Fe and Cr and amplifies their interaction in comparison with a perfect vacancy-free Fe-Cr alloy.

Finally, we studied structural relaxations of atoms from ideal lattice positions around a vacancy. In the nearest neighbour position, both Fe and Cr atoms tend to move slightly towards the vacancy, but the displacements are relatively small. For the case of iron atoms only, their displacement from the first shell to the vacancy is 0.085 Å. In the second shell, Fe atoms move away from the vacancy by 0.028 Å. This is in agreement with calculations of the Kanzaki forces and defect dipole tensors [44], where inward relaxation was found for the first nearest neighbours and outward relaxation for the second nearest neighbours. Similar behaviour, but with smaller relaxations, is observed for chromium atoms. If a single Fe atom in the first shell is replaced by a Cr atom, in the relaxed position this atom is shifted 0.056 Å towards the vacancy. Iron atoms in the first coordination shell also relax their position in the direction towards the vacancy, the displacements are in the interval from 0.072 Å to 0.095 Å. For a single chromium atom in the second nearest neighbour position, the displacement is 0.033 Å away from the vacancy (Fe atoms in the second coordination shell also relax away from the vacancy by between 0.025 Å to 0.031 Å). With increasing number of Cr atoms in the first shell, their relaxation becomes smaller and ultimately changes sign for the case of eight Cr atoms, where these atoms relax away from the vacancy (by 0.007 Å only). This is related to the repulsive magnetic interaction between chromium atoms discussed above.

Overall, we found that the bcc structure both with and without Cr atoms is distorted only slightly in the vicinity of a vacancy.

**Divacancy-chromium clusters**

If a vacancy cluster contains more than just a single vacancy, the number of possible configurations of Cr atoms around the vacancy cluster becomes prohibitively high for *ab initio* calculations, and it proves necessary to restrict ourselves to the treatment of only a selected number of cases. We have chosen to study vacancy clusters where the two vacancies are in the nearest and second nearest neighbour positions, because the binding energy is the largest in these two cases (see Table 1). Only clusters with one and two chromium atoms were considered, with a constraint that the largest separation distance between vacancies or the chromium atoms in the cluster is less than half of the simulation cell size (i.e., less than the sixth nearest neighbour).

For the two vacancies in the nearest neighbour configuration, three possible cluster structures studied are shown in Figure 7 (a-c). We found that the binding energy of a cluster consisting of two vacancies and one chromium atom is almost independent on the position of the chromium atom with respect to vacancies in all the configurations shown in Figure 7 (a-c). For configuration 7 (a), the calculated binding energy is 0.278 eV, whereas for configurations shown in Figures 7 (b) and 7 (c), the binding energy is 0.289 eV. Hence, the addition of a single Cr atom increases the binding energy of the cluster by more than 100 meV, from 0.175 eV (see Table 1) to 0.289 eV.

In configurations where the vacancies are in the second nearest neighbour position (Figure 7 (d-e)), there are only two symmetrically non-equivalent configurations for a single Cr atom and the binding energy strongly depends on its position. In this case, a chromium atom prefers the symmetric position where it is the nearest neighbour of both vacancies (Figure 7 (d)), with the total binding energy increasing from 0.234 eV (Table 1) to 0.333 eV. If the Cr atom is in a non-symmetric position where it is the nearest neighbour of one vacancy and is the fourth nearest neighbour of the second vacancy (Figure 7 (e)), the binding energy is close to 0.236 eV.

In configurations containing two vacancies and two Cr atoms, 12 different configurations with vacancies being nearest neighbours and 7 configurations with them being second nearest neighbours were investigated. It is instructive to look at configurations characterised by the highest and the lowest binding energies. In the case of vacancies in the nearest neighbour configuration, the lowest binding energy is -0.014 eV (i.e., the cluster is marginally unstable). This corresponds to the case where the two Cr atoms are the nearest neighbours of each other (Figure 8 (a)). The highest binding energy of 0.348 eV was found in the case of Cr atoms in the fourth nearest neighbour positions (Figure 8 (b)). This configuration is not the only one that has a high binding energy: altogether we have found eight structures where the binding energy is in the range from 0.31 to 0.35 eV. In all of them, chromium atoms are either in the third, or the fourth, or the fifth neighbour position, confirming the qualitative picture of bonding established above for the case of single vacancy. Similarly, for the vacancies in the second nearest neighbour position, the lowest binding energy of 0.188 eV (which is lower than the binding energy between two vacancies only) was found where Cr atoms are in the second nearest neighbour position (Figure 8 (c)), and the highest (0.409 eV) where they are third nearest neighbours of each other (Figure 8 (d)). In general, the binding energy falls into two bands: 0.18-0.22 eV where Cr atoms are second nearest neighbours and 0.32-0.41 eV where they are further apart. This, together with a similar observation applied to the case of vacancies in the nearest neighbour position, suggests that it should be possible to describe the binding energy of vacancy-Cr clusters in terms of a cluster expansion-type model. Comparing single vacancy and di-vacancy clusters, we found that for the former, the ratio of the binding energy to the number of Cr atoms is the highest for the case of a single chromium atom in the nearest neighbour position (0.094 eV, see Table 2). For the di-vacancy clusters studied, the ratio of the binding energy to the number of Cr atoms is the highest for the cluster shown in Figure 8 (d) and is 0.204 eV per atom, i.e. increased more than two times compared to the single vacancy case.

Similarly to the case of a single vacancy, there is a clear correlation between the average magnetic moment of chromium atoms and the binding energy, as shown in Figure 9. Larger distances between Cr atoms result in less strong repulsive Cr-Cr magnetic interactions. As a result, magnetic moments increase, and the positive Fe-Cr magnetic coupling gives rise to higher value binding energies. As in the case of a single vacancy, displacements of Cr and Fe atoms from their ideal lattice positions are small in comparison with the lattice parameter.

**Discussion**

In this work, we performed an *ab initio* investigation of small vacancy and vacancy-chromium clusters in bcc iron. The objective of this study was to investigate the possibility of clustering of Cr atoms around vacancies in Fe-Cr alloys at relatively low chromium concentrations (below the thermodynamic solubility limit of 10-11 at. %, where Cr clustering begins in a perfect bcc alloy). We found that a single vacancy attracts up to four Cr atoms in the most energetically favourable configuration. Overall, the binding energy of a single vacancy cluster with 1 to 8 Cr atoms can be approximated by a linear fit as a function of the number of the second, third and fifth nearest Cr-Cr neighbours. Chromium atoms around a vacancy prefer to be in the fifth and the third nearest neighbour position with respect to each other, while being in the second nearest neighbour lowers the binding energy. The magnetic origin of this trend in the binding energy is confirmed by a correlation between the average value of magnetic moment of Cr atoms and the binding energy. Similar trend is found also for the divacancy-Cr clusters studied here. Magnetic interaction in the presence of a vacancy or a vacancy cluster is amplified compared to the case of a perfect Fe-Cr alloy because of the increase of Fe and Cr magnetic moments.

Our results confirm that single vacancies, as well as divacancy clusters, can attract chromium atoms, forming small localised volumes of high Cr concentration that cannot be detected using the currently available experimental means. Further investigation of this topic should aim at the development of a cluster expansion type model for the binding energy that would allow extending the scale of simulations to sizes that are currently too large for *ab initio* analysis. Also, for the assessment of stability of Cr-vacancy clusters, it is necessary to evaluate not only the binding energy, but also the height of the barrier that has to be overcome to dissociate the cluster. Hence, calculations of barrier heights and diffusion pathways for vacancy-Cr and vacancy-Fe exchanges in these clusters are required. These calculations, together with large-scale model for the binding energy, should enable carrying out kinetic Monte Carlo simulations of the evolution of vacancy-chromium clusters in the Fe matrix. Finally, even though we did not find large displacements of atoms from their ideal crystallographic positions, this possibility cannot be excluded in the limit of a large number of vacancies in a cluster. For larger vacancy and vacancy-Cr clusters, it is natural to expect large lattice deformations that could in turn result in structural transformation away from the body centered cubic lattice, at least on small scales. Transformation of clusters of self-interstitial atom defects in pure Fe into clusters of the C15 phase was predicted recently in Ref. [45]. It was also shown that microstructure may change significantly in alloys where the mismatch between atomic sizes is small [46]. In our case, vacancies and Cr atoms introduce substantial strains in ideal bcc Fe lattice, so an investigation of the possibility of structural and microstructural changes associated with large vacancy and vacancy-Cr clusters warrants attention.

**Acknowledgments**

This work was part-funded by the EuroFusion Consortium, and has received funding from Euratom research and training programme 2014-2018 under grant agreement number No. 633053, and funding from the RCUK Energy Programme (Grant Number EP/I501045). The views and opinions expressed herein do not necessarily reflect those of the European Commission. To obtain further information on the data and models underlying this paper please contact PublicationsManager@ccfe.ac.uk. This work was also part-funded by the United Kingdom Engineering and Physical Sciences Research Council via a programme grant EP/G050031. MYL and DNM would like to acknowledge the International Fusion Energy Research Centre (IFERC) for the provision of a supercomputer (Helios) at the Computational Simulation Centre (CSC) in Rokkasho (Japan).

| Number of vacancies | This work | [12] | [23] | [19] | [18] | [11] | [10] |
|---|---|---|---|---|---|---|---|
| 2 (NN) | 0.175 (0.140) | 0.184 | 0.16 | 0.17 | 0.06 | 0.044 (0.224) | 0.131 |
| 2 (2NN) | 0.234 (0.224) | 0.194 | 0.23 | 0.23 | 0.15 | 0.054 (0.150) | 0.195 |
| 2 (3NN) | 0.045 (-0.001) | | | | | | |
| 2 (4NN) | 0.055 | | | | | | |
| 3 | 0.688 | 0.66 | | | | 0.241 (0.670) | 0.489 |
| 4 (tetragonal) | 1.478 | 1.31 | | | | 0.660 (1.264) | 1.023 |
| 4 (square) | 1.116 (0.975) | 0.59 | | | | 0.034 (0.488) | 0.75 |

Table 1. Binding energies of clusters of 2, 3, and 4 vacancies (eV) obtained in the present study and compared with calculations performed by others. Values in parentheses obtained in this work are calculated using larger simulation boxes containing 5×5×5 unit cells. In the paper by Masuda [11], the values in parentheses were obtained for unrelaxed atomic configurations.

| Vacancy – Cr distance | This work |
|---|---|
| NN | 0.094 |
| 2NN | 0.053 |
| 3NN | 0.043 |
| 4NN | -0.010 |
| 5NN | -0.001 |

Table 2. Binding energies for vacancy-Cr pairs (eV) as functions of separation.

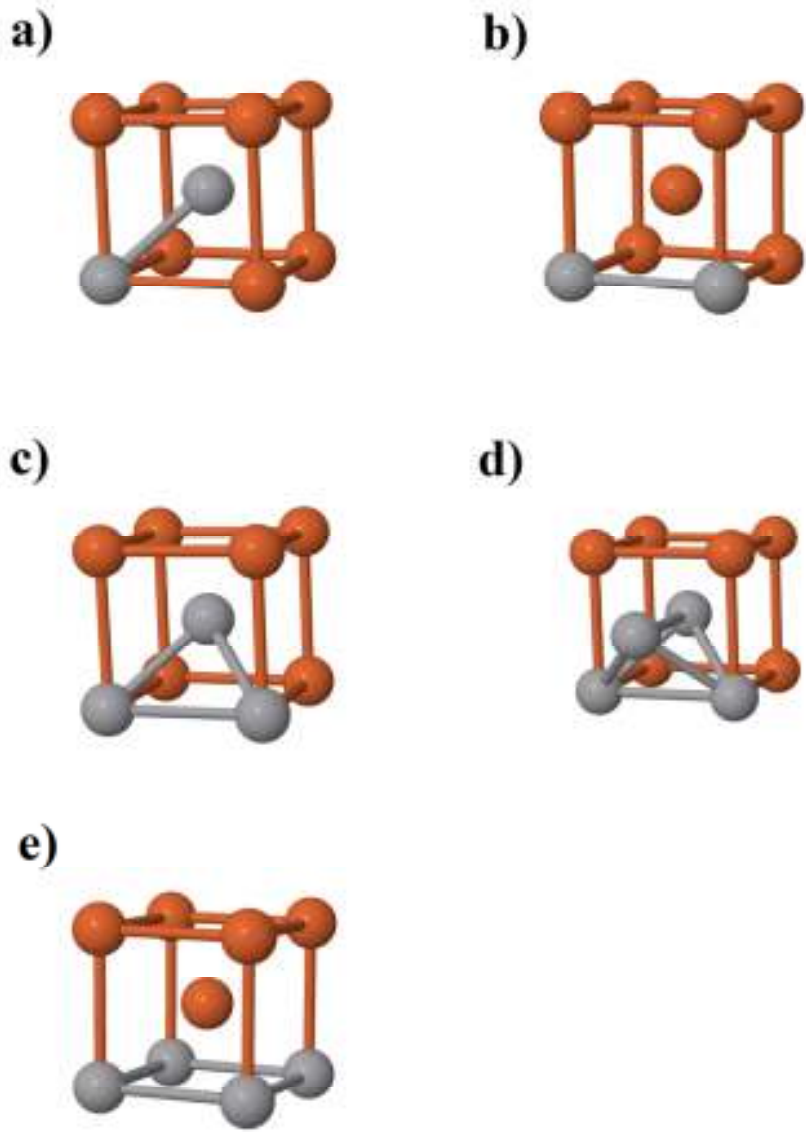

Figure 1. Configurations containing two (NN (a) and 2NN (b)), three (c) and four (d,e) vacancies investigated by *ab initio* calculations. Fe atoms are shown as light brown spheres, vacancies as grey spheres.

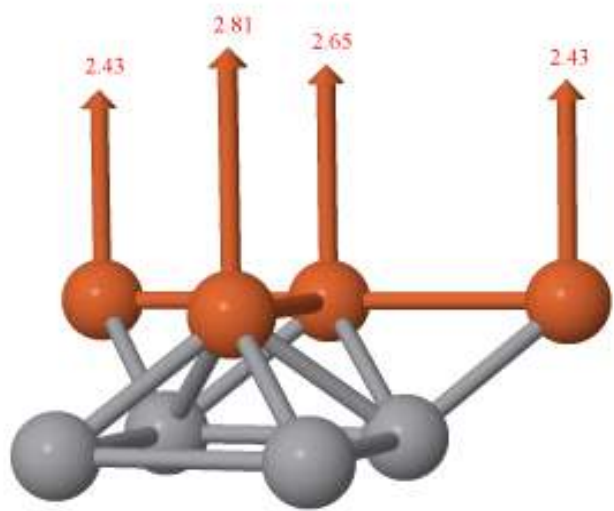

Figure 2. Magnetic moments ($\mu_B$) of Fe atoms in the vicinity of a four vacancy cluster adopting a "square" configuration.

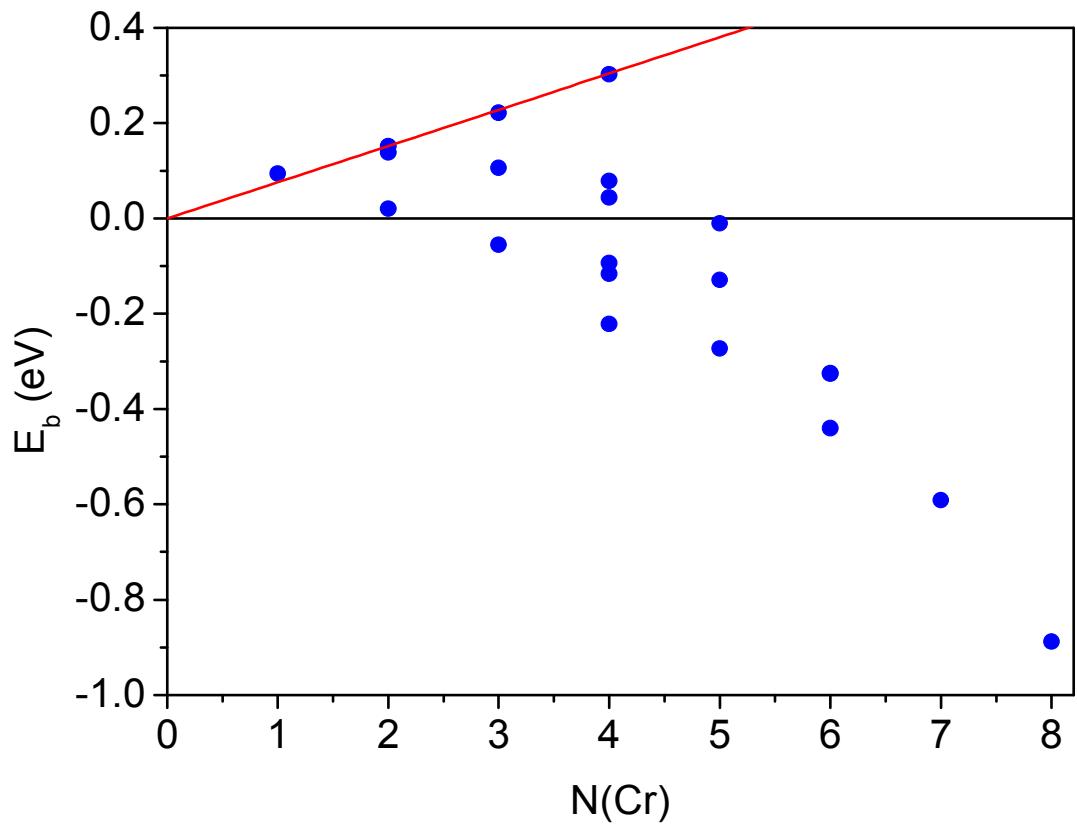

Figure 3. Binding energy $E_b$ of vacancy-chromium clusters as a function of the number of Cr atoms in the first coordination shell around a vacancy (in eV).

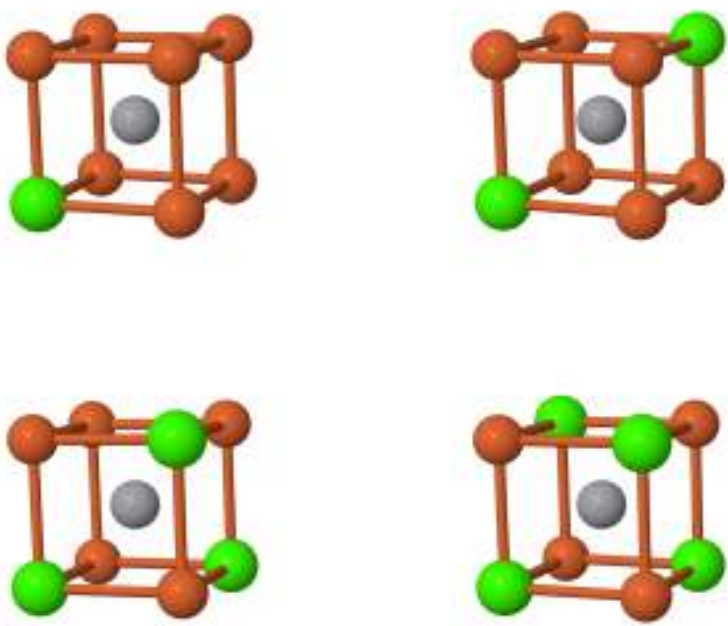

Figure 4. Configurations with the highest binding energy, containing from 1 to 4 Cr atoms (green spheres) around a vacancy (grey sphere).

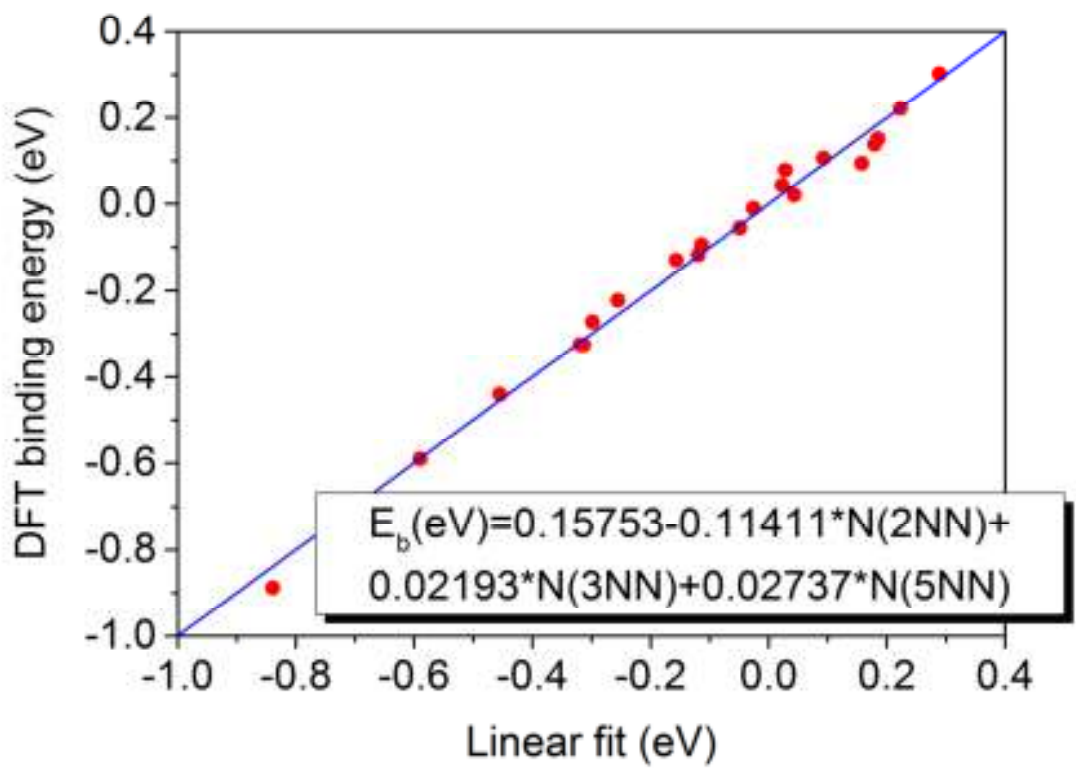

Figure 5. Plot illustrating agreement between DFT binding energy data on single vacancy-Cr clusters and a linear fit with respect to the number of the nearest, third and fifth nearest Cr-Cr neighbours.

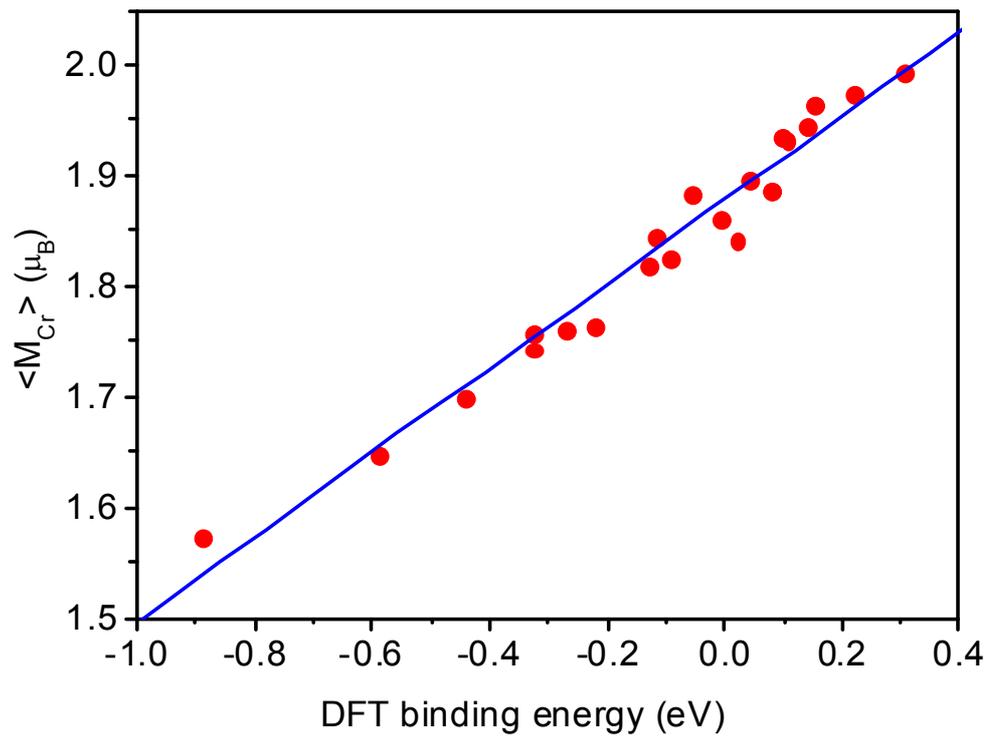

Figure 6. Average value of magnetic moment of Cr atoms ($\mu_B$) around a single vacancy as a function of the binding energy of the cluster (eV). The straight line is a linear fit $\langle M_{Cr} \rangle = 1.879 + 0.394 * E_b$.

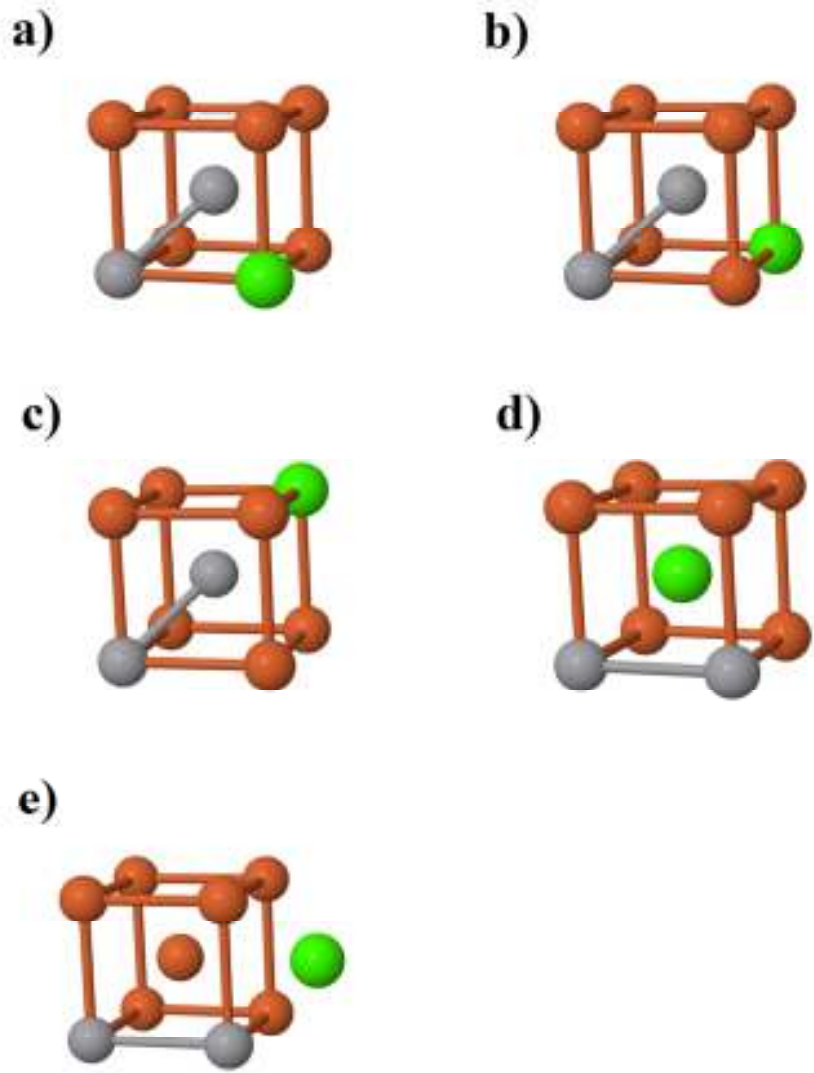

Figure 7. Configurations containing two vacancies in the nearest (a-c) or second (d-e) nearest neighbour position also containing a single Cr atom.

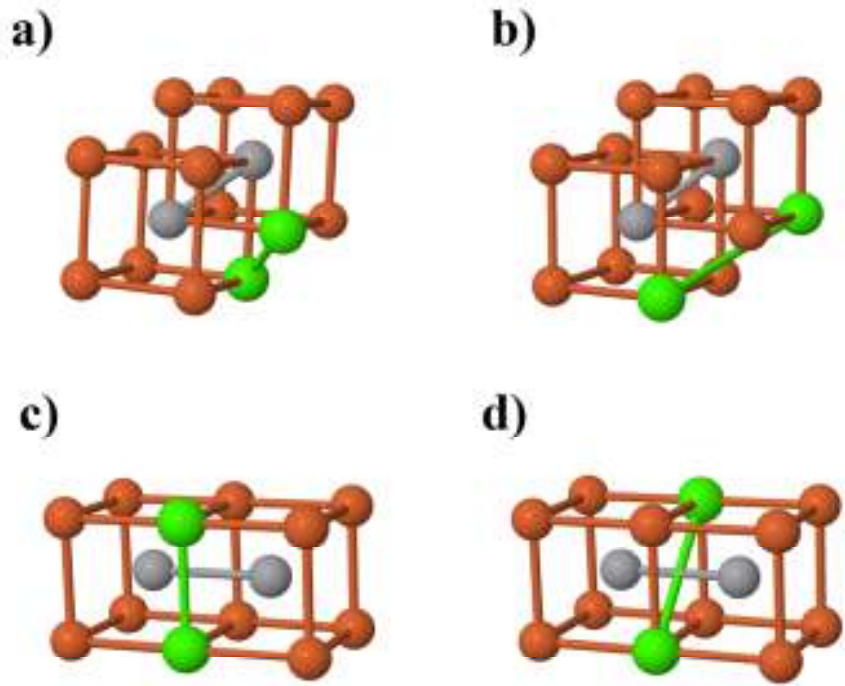

Figure 8. Lowest (a, c) and highest (b, d) energy configurations containing two vacancies and two Cr atoms. Vacancies are either in the nearest (a, b), or second nearest (c,d) neighbour position with respect to each other.

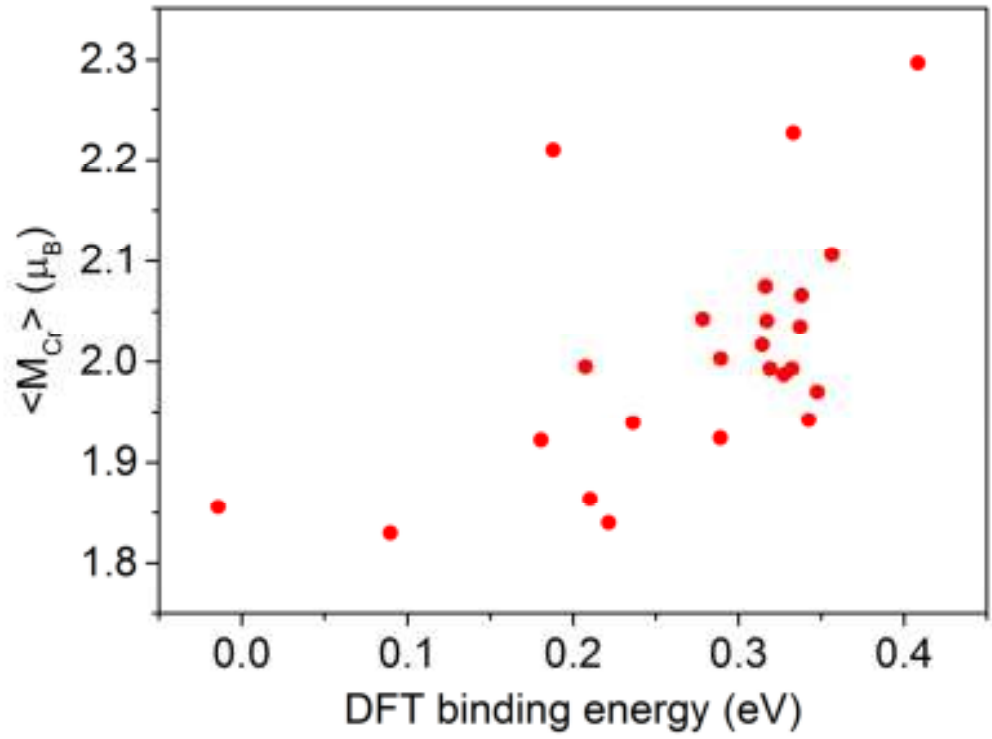

Figure 9. Average magnetic moment of a Cr atom ($\mu_B$) in the vicinity of two vacancies as a function of the binding energy of the cluster (eV).